\documentclass[namedreferences]{solarphysics}
\usepackage[optionalrh,solaenum]{spr-sola-addons} 
\usepackage{graphicx}                    
\usepackage{color}                       
\usepackage{url}                         

\begin{document}

\begin{article}

\begin{opening}

\title{Study of a Prominence Eruption using SWAP/PROBA2 and EUVI/STEREO Data}

%
 \author{M.~\surname{Mierla}$^{1,2,3}$\sep
        D.B.~\surname{Seaton}$^{2}$\sep
        D.~\surname{Berghmans}$^{2}$\sep 
        I.~\surname{Chifu}$^{4,5}$\sep
        A.~\surname{De Groof}$^{6}$\sep 
        B.~\surname{Inhester}$^{4}$\sep
        L.~\surname{Rodriguez}$^{2}$\sep 
        G.~\surname{Stenborg}$^{7}$\sep        
        A.N.~\surname{Zhukov}$^{2,8}$
       }

%
\runningauthor{M. Mierla et al.}
\runningtitle{Prominence Eruption}

%
\institute{$^{1}$ Institute of Geodynamics of the Romanian Academy, Bucharest, Romania
                     email: \url{marilena@geodin.ro}
             $^{2}$ Solar-Terrestrial Center of Excellence - SIDC, Royal Observatory of Belgium, Brussels, Belgium
             $^{3}$ Research Center for Atomic Physics and Astrophysics,
                  Faculty of Physics, University of Bucharest, Romania
             $^{4}$ Max-Planck Institute for Solar System Research, Katlenburg-Lindau, Germany
             $^{5}$ Astronomical Institute of the Romanian Academy, Bucharest, Romania
             $^{6}$ European Space Agency / Royal Observatory of Belgium, Brussels, Belgium
             $^{7}$ School of Physics, Astronomy and Computational Sciences, George Mason University, Fairfax, VA, USA
             $^{8}$ Skobeltsyn Institute of Nuclear Physics, Moscow State University, Moscow, Russia            
             }

\begin{abstract}
Observations of the early rise and propagation phases of solar eruptive prominences can provide clues about the forces acting on them through the behavior of their acceleration with height.
We have analyzed such an event, observed on 13 April 2010 by SWAP on PROBA2 and EUVI on STEREO. A feature at the top of the erupting prominence was identified and tracked in images from the three spacecraft. The triangulation technique was used to derive the true direction of propagation of this feature. The reconstructed points were fitted with two mathematical models: i) a power-law polynomial function and ii) a cubic smoothing spline, in order to derive the accelerations. The first model is characterized by five degrees of freedom while the second one is characterized by ten degrees of freedom. The results show that the acceleration increases smoothly and it is continuously increasing with height. We conclude that the prominence is not accelerated immediately by local reconnection but rather is swept away as part of a large-scale relaxation of the coronal magnetic field. 
\end{abstract}

%

\end{opening}

%
 \section{Introduction}
Prominences (or filaments, when seen on the solar disk) are known to mark the launch site of coronal mass ejections (CMEs), which may sometimes arrive at the Earth and trigger severe geomagnetic storms. This is why there is an extensive effort to study these solar phenomena (\textit{e.g.} the reviews by \opencite{Labrosse10}; \opencite{Mackay10}; \opencite{Bemporad11}). In general, an eruption can be described by three phases: i) the initiation phase; ii) the main acceleration phase and iii) the propagation phase ( \textit{e.g.} \opencite{Zhang06}). The explanation of these phases are given below.

i) In order to initiate the first phase of the eruption, different triggering mechanisms are invoked (see the review by \opencite{Chen11}): tether-cutting or flux cancellation; shearing motions; magnetic breakout; emerging flux triggering; flux injection triggering; instability and catastrophe related triggering (kink instability; torus instability; catastrophe); hybrid mechanisms; as well as other mechanisms (mass drainage, sympathetic effect, solar wind).

By fitting the height--time diagram of the tracked features in the solar corona with a power-law polynomial and by comparing it with the results derived from various numerical simulations, one may get clues to the destabilization mechanisms involved (\textit{e.g.} \opencite{Torok05}; \opencite{Williams05}).

Different power-law exponents were found for different eruption mechanisms: 
\renewcommand{\labelenumi}{\alph{enumi})}
\begin{enumerate}
\item a power-law exponent ($m = 2.5$) fits the catastrophe model scenario best \cite{Priest02}; 
\item a parabolic profile ($m = 2$) characterizes the CME rise in a breakout model simulation well \cite{Lynch04}; 
\item an exponent of 3 is closer to the torus instability scenario where a sufficiently large initial perturbation is applied \cite{Schrijver08}. 
\item an exponential rise has been observed for a three-dimensional flux rope subject to a helical kink instability \cite{Torok04,Torok05}. 
\item the exponential rise is also valid for the torus (expansion) instability, which starts as a $\sinh(t)$ function that is similar to a pure exponential \cite{Kliem06}. 
\end{enumerate}

ii) The acceleration phase occurs when the erupting prominence experiences a very rapid increase in its velocity. Usually, it is assumed that this phase is triggered either by magnetic reconnection taking place at the site of the eruption or by an ideal MHD instability. \inlinecite{Zhang01} found that the acceleration phase of an erupting prominence takes place simultaneously with the impulsive phase of the associated flares, indicating that the main magnetic reconnection plays an important role in accelerating the filaments. In some events the rapid acceleration phase is not observed, meaning that some factors other than reconnection may also contribute to the acceleration of CMEs, \textit{e.g.}, ideal MHD instabilities \cite{Amari04,Fan07}. This may be the reason why these events undergo continuous acceleration before and after the impulsive phase (\textit{e.g.}, \opencite{Kahler88}).

 iii) At higher altitudes, after the impulsive acceleration phase, most erupting filaments rise with an almost constant velocity (\textit{e.g.}, \opencite{Kahler88}; \opencite{Sterling05}; \opencite{Joshi07}). This usually defines the propagation phase. 

The main acceleration and the propagation phases can be understood if we consider the forces acting on an erupting prominence during each of them: basically the Lorentz force, the gravitational force, and the drag due to the ambient solar wind. \inlinecite{Chen03} and \inlinecite{Chen06} have identified the main acceleration phase as the interval during which Lorentz forces are dominant, while during the gradual propagation phase the Lorentz force is comparable with the other two forces. 

Usually, the eruptions are associated at later times with CMEs, while at the initial moments they are referred to as precursors. To avoid confusion, we adopt the definitions of \inlinecite{Chen11}: ``the progenitor is the unstable or metastable coronal structure that would be the source of a CME'' (in the case of the event discussed in this article the progenitor is the prominence) and the precursor is ``an observable signature associated with the initiation of the CME, \textit{i.e.}, before its main acceleration phase'' (in the case of the event discussed in this article the precursor is the slow rise of the prominence). 
In the case of CMEs associated with prominence eruptions such an observational precursor is the slow rise of prominences \cite{Filippov08}. Other precursors associated with filament eruptions (\textit{e.g.} \opencite{Chen11}) may be: filament darkening and widening \cite{Martin80}; long-term period filament/prominence oscillations \cite{Chen08}; or reconnection-favoring emerging flux (\textit{e.g.} \opencite{Feynman95}). All of these precursors are signatures associated with the moment of initiation of the CME, except the emerging flux which appears well before the onset of CMEs. However, as pointed out by \inlinecite{Chen11}, precursors are neither a necessary nor sufficient condition for the onset of an eruption.

In this article we analyze the two first phases of a prominence eruption that occurred on 13 April 2010. We aim to give a more systematic definition of the different phases explained above. The event was observed by EUVI on STEREO and SWAP on PROBA2. The event was also the progenitor of a CME observed later in LASCO and STEREO/COR white-light images.

STEREO was launched in October 2006 and continuously observes the Sun from two viewpoints at approximately the same distance from the sun as the Earth; the angle between them increasing at a rate of 42 degrees per year. PROBA2 was launched in November 2009 into a Sun-synchronous orbit around the Earth. It is used in this study as the third eye on the Sun. Using STEREO and PROBA2 observations together, we employed the triangulation technique in order to derive the true direction of propagation and the true speed of the top of the prominence. The data were fitted with two mathematical models in order to derive the acceleration of the prominence. First, a polynomial function, with five degrees of freedom: $h=h_0+v_0 \cdot (t-t_0)+a_0 \cdot (t-t_0)^m$ (see \opencite{Alexander02}) was fitted to the 3D reconstructed points. Alternatively, a spline smoothing procedure (which implies more degrees of freedom) was applied to the data. From the above-mentioned studies it is not clear when each of the eruption phases starts. By applying the two independent methods, we aim to determine the exact division point between the two phases observed during our event: the initiation phase and the acceleration phase. Comparison of the results derived from the two methods are described and discussed. 

\section{Observations}
\begin{figure*}[!th]
  \centering
  \includegraphics[width=.3\textwidth,type=eps,ext=.eps,read=.eps]{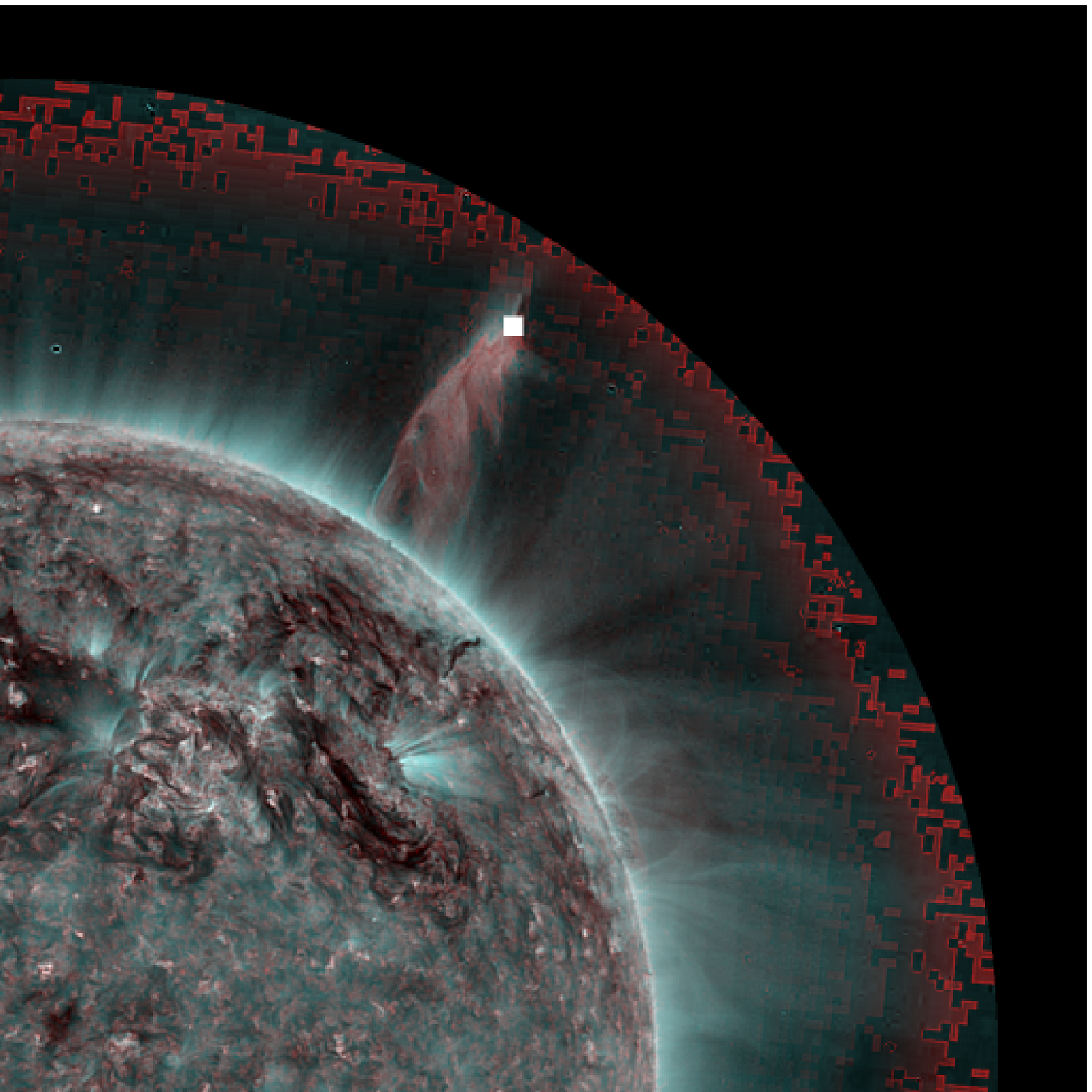}
  \includegraphics[width=.3\textwidth,type=eps,ext=.eps,read=.eps]{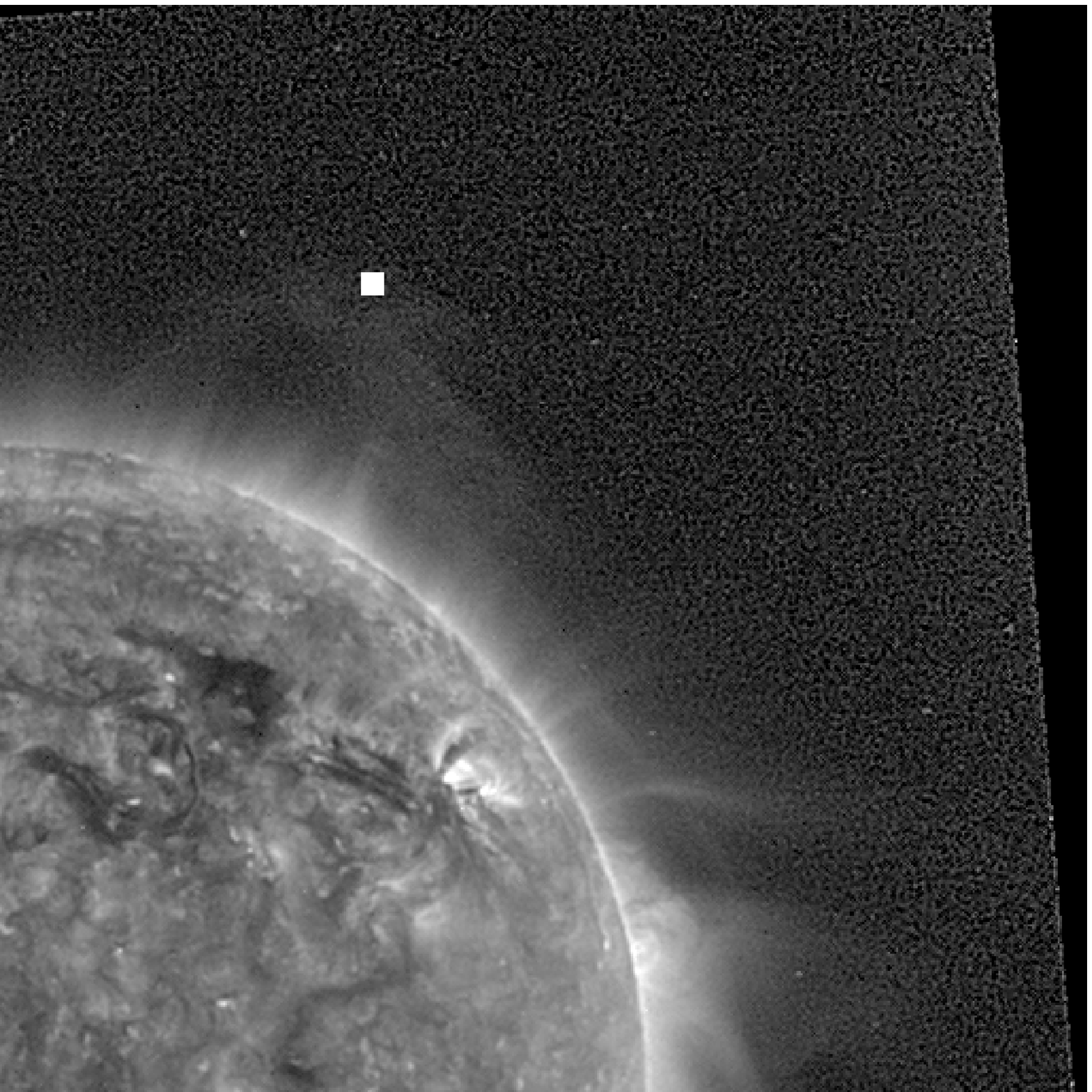}
  \includegraphics[width=.3\textwidth,type=eps,ext=.eps,read=.eps]{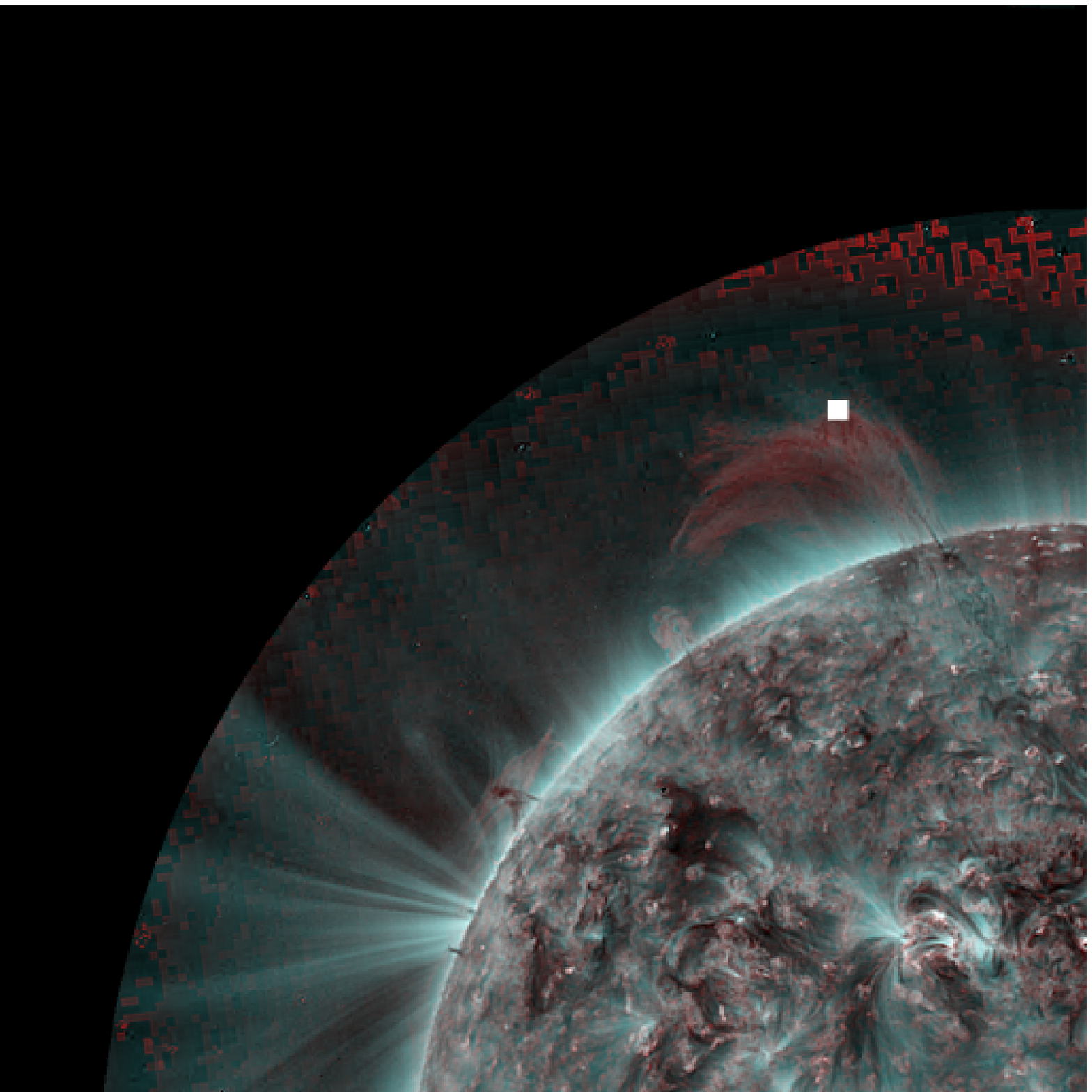}
  \caption{Left and right panels: Composite of STEREO EUVI wavelet-enhanced images (left: S/C B; right: S/C A) in 171~\AA\ (blue) and 304~\AA\ (red) showing the prominence material at 08:16:15 UT and its counterpart in a higher temperature regime at 08:14:00 UT. Middle panel: Prominence observed by SWAP 174~\AA\, at around 08:15~UT on 13~April~2010. The images were processed using the wavelet technique. The center of the white square shows the location of the 3D reconstructed feature (TOP-3D) back projected on the 2D images.}
  \label{swapeuvi304}
\end{figure*}

On 13 April 2010 at about 06:00 UT, a huge, bright prominence erupted from the northwest quadrant of the Sun as seen from the Earth.  We obtained observations of this eruption using the \emph{Sun-Watcher with Active Pixel System and Image Processing} (SWAP) \cite{Berghmans06,Seaton12} onboard the Project for On-Board Autonomy 2 (PROBA2) spacecraft in low-earth orbit and the \emph{EUV Imagers} (EUVI-A, -B) \cite{Howard08} onboard the \emph{Solar-Terrestrial Relations Observatory} (STEREO-A, -B) spacecraft. At the time of the eruption, the separation angle between the two STEREO spacecraft was about 139~degrees, while the separation angle between STEREO-A and Earth was about 68~degrees. Both spacecraft were essentially in the ecliptic plane at that time. Figure~\ref{swapeuvi304} shows the eruption as it appeared on PROBA2 in Earth-orbit and on the two STEREO spacecraft.

The prominence appeared brightest in EUVI-B's 304~\AA\ passband, and we first determined the location of the top of the prominence in these images. (We will refer to this feature as TOP-B -- see the white square, Figure~\ref{swapeuvi304}, left panel). We then combined these images with EUVI-A images to reconstruct the location of this feature in three-dimensional space (we refer to this as the first reconstruction). To do this we employed the triangulation technique, based on the epipolar constraint \cite{Inhester06}, which is implemented in the \textsf{scc\_measure.pro} program in \textsf{SolarSoft}. The standard triangulation technique can most often be used only when the angular separation between the vantage points is not too large \cite{Inhester06,Rodriguez09}. Special care is needed when working with large angular separations, which is the reason that we chose to look only at the top of the prominence where there is little ambiguity about the feature's location. Note that when we say feature, we refer only to the single point at the top of the prominence.

In images from EUVI-A, the top of the prominence is more broad, and sometimes more than one feature intersects the projected line-of-sight (LOS) corresponding to TOP-B. In those cases we manually selected what we determined was the most probable corresponding feature from EUVI-A images to make the reconstruction. We made this selection by comparing EUVI-A and -B movies side-by-side. 

We computed the uncertainty in the three-dimensional position of the top of the prominence by measuring the distance between the feature we selected in the EUVI-A images and the most distant feature that intersected the corresponding epipolar line. We then determined the three-dimensional uncertainty by converting these locations into the corresponding height, latitude, and longitude in three-dimensional space. It is worth pointing out that, because of the projection between points-of-view for these images, this technique tends to produce accurate uncertainties for the depth (that is, longitude) of each feature that we identified, but overestimates the uncertainty in latitude and height above the solar surface. Although these effects are relatively small for our height--time curve, they are grossly enlarged if the curve is differentiated to obtain speed and acceleration.

We applied the same reconstruction technique to EUVI-B~304~\AA\ and SWAP 174~\AA\ images. (We refer to this as the second reconstruction). We use EUVI-B~304~\AA\ images instead of EUVI-B~171~\AA\ because the cadence of the 171~\AA\ observations is low (one image every two hours). The appearance of the prominence in SWAP~174~\AA\ images is somewhat different from its appearance in the cooler lines that appear in 304~\AA\ images, but the top of the prominence is nonetheless easy to identify as a bright, persistent feature through almost all of the images in the time series. Later in the progression of the eruption, this feature appeared bright against the dark background of the off-limb corona, but the top of the prominence was still easy to identify (see also the left and right panels of Figure 1). 

Before the eruption, a part of the prominence is clearly seen in absorption in the 174~\AA\ SWAP images. Other parts of the prominence can also be seen in emission, although it is not easy to distinguish them from each other due to superposition with bright coronal structures along the line of sight. However, as the eruption unfolds, both emitting and absorbing parts of the prominence begin to rise. It is also possible that, during the eruption, some parts of the prominence are heated (as discussed by \inlinecite{Filippov02}, for example) and the prominence becomes visible mostly in emission. In the absence of spectroscopic data, it is difficult to determine the evolution of the prominence temperature and, thus, the reasons that it is clearly visible in both the 171 and 304~\AA\ bandpasses (see STEREO/SECCHI images, left and right panels of Figure 1).

That we observe the prominence in emission in both bandpasses suggests that it appears not in the most prominent lines of SWAP's 174~\AA\ bandpass (Fe~{\sc ix} and Fe~{\sc x}, formed around 0.8 -- 1~MK), but rather in weaker lines that form at a lower temperature. Although the 304~\AA\ bandpass does contain lines such as Si~{\sc xi} that form at high temperatures, the appearance of the prominence here is consistent with other observations of erupting prominences that were determined to have temperatures corresponding to the chromosphere and transition region (around 0.01 -- 0.3~MK) using spectroscopic observations (\textit{e.g.} \opencite{Gunar11}). That the prominence is optically thick in 304~\AA\ observations further suggests that the material we see in the EUVI observations corresponds to emission from the cooler He~{\sc ii} line and, correspondingly, in SWAP observations to the two O~{\sc vi} lines at 173~\AA\ \cite{DelZanna03} that form close to 0.3~MK. However, we cannot rule out the possibility that some non-equilibrium heating processes that may occur during the eruption could be contributing to the appearance of this prominence in emission around 171~\AA\ (\textit{e.g.} \opencite{Filippov02}).

Because the cadence of SWAP images was about two minutes, while the cadence of EUVI-B images was five minutes, we occasionally encountered situations where no single SWAP image could be matched directly to an EUVI image. In this case, we used pairs of SWAP images to perform the reconstruction, resulting in two reconstructed points for a single EUVI-B feature. However, we restricted the maximum time difference between images in the reconstruction to 65 seconds. Since the eruption had a maximum speed of about 60~km s$^{-1}$ early in its rise, the differences were about $6\times10^{-3}$~solar-radii. Since this error is smaller than the inherent uncertainties in our reconstruction procedure (which ranged from about $7\times10^{-3}$ to $6\times10^{-2}$~solar-radii) we conclude that these differences in time have a negligible effect on our reconstruction of the height of the prominence. An example of the reconstructed feature, back-projected on the three images, is shown in Figure 1 as the center of a white square.

\begin{figure*}[!th]
  \centering
  \includegraphics[width=.9\textwidth,type=eps,ext=.eps,read=.eps]{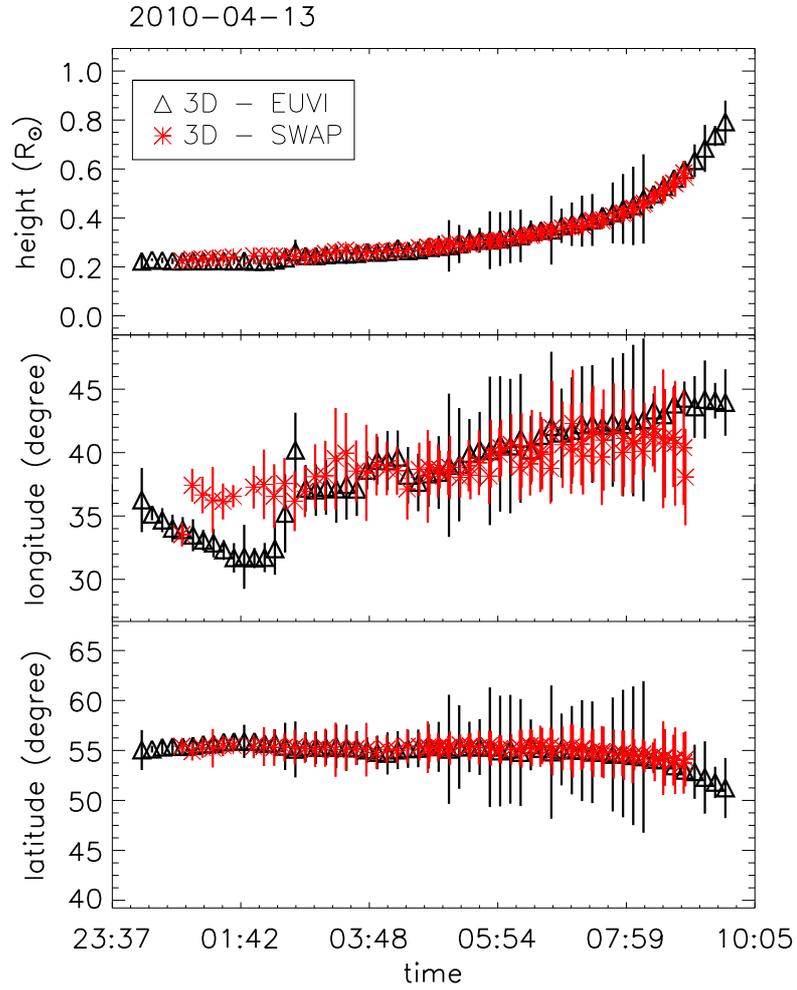}
  \caption{Height above the surface, longitude, and latitude of the top of the prominence \textit{versus} time. The coordinates were derived using the 3D triangulation technique applied on EUVI-A and -B 304~\AA\ images (first reconstruction, black triangles) and EUVI-B 304~\AA\ and SWAP 174~\AA\ images (second reconstruction, red asterisks).}
  \label{ht3d304}
\end{figure*}

In this study we focus on the early part of the evolution of the eruption, as we are interesting in determining the initiation mechanism. Figure~\ref{ht3d304}, which shows a height--time diagram for the reconstructed feature (which we refer to as TOP-3D), reveals that the feature rose almost imperceptibly in the first five or six hours of its evolution, and only began its rise at about 06:00 UT. Throughout the eruption, the latitude and longitude remain relatively constant (within the error limits). The longitude increased from 35 to 45 degrees from the beginning to the end of the observed period, and the latitude decreased slightly from 55 to 50 degrees starting at 8:35~UT. Note that the calculated longitude includes a larger random error than the calculated latitude because longitude (or depth) is the most sensitive to reconstruction errors (see \opencite{Inhester06}). The decrease of the latitude indicates a commonly observed deflection of the erupting structure towards the equator (\textit{e.g.} \opencite{Kilpua09}). It is probably due to the magnetic-force imbalance in the latitudinal direction as indicated by recent MHD simulations \cite{Zuccarello12}.

\section{Analysis}
We employed two different approaches in our analysis of the prominence eruption. First, the three-dimensional reconstructed heights were fitted with a polynomial function.  Second, we obtained a smooth function using a spline smoothing procedure (described below) in order to derive the velocity and acceleration as a function of time. In the following section we describe each method in detail.

\subsection{Polynomial Fitting}
 Several authors have used polynomial fits to study the acceleration of CMEs in the low corona (\textit{e.g.} \opencite{Alexander02}). Here we fit the three-dimensional reconstructed heights with the same function as \inlinecite{Schrijver08} used in their analysis of the rise profile of a prominence observed by TRACE:

\begin{equation}
h(t)=h_0 + v_0 \left( t-t_0 \right) + a_0 \left( t-t_0 \right)^{m},
\end{equation}
where $h_0$ is the initial height of the prominence with units of km, $v_0$ is the initial velocity perturbation due to destabilization during the initiation phase and has units of km s$^{-1}$, $a_0$ is a parameter that represents the rate of change of acceleration and has units $\mathrm{km} \mathrm{s}^{-m}$, and $t_{0}$ is an initial displacement in time before which the acceleration is negligible. It is worth pointing out that the rise modeled by \inlinecite{Schrijver08} covered a much shorter duration (only a few minutes) and, because of TRACE's small field-of-view, was restricted to low heights in the corona compared to our event.

\begin{figure*}[!th]
      \centering
      \includegraphics[width=.49\textwidth,type=eps,ext=.eps,read=.eps]{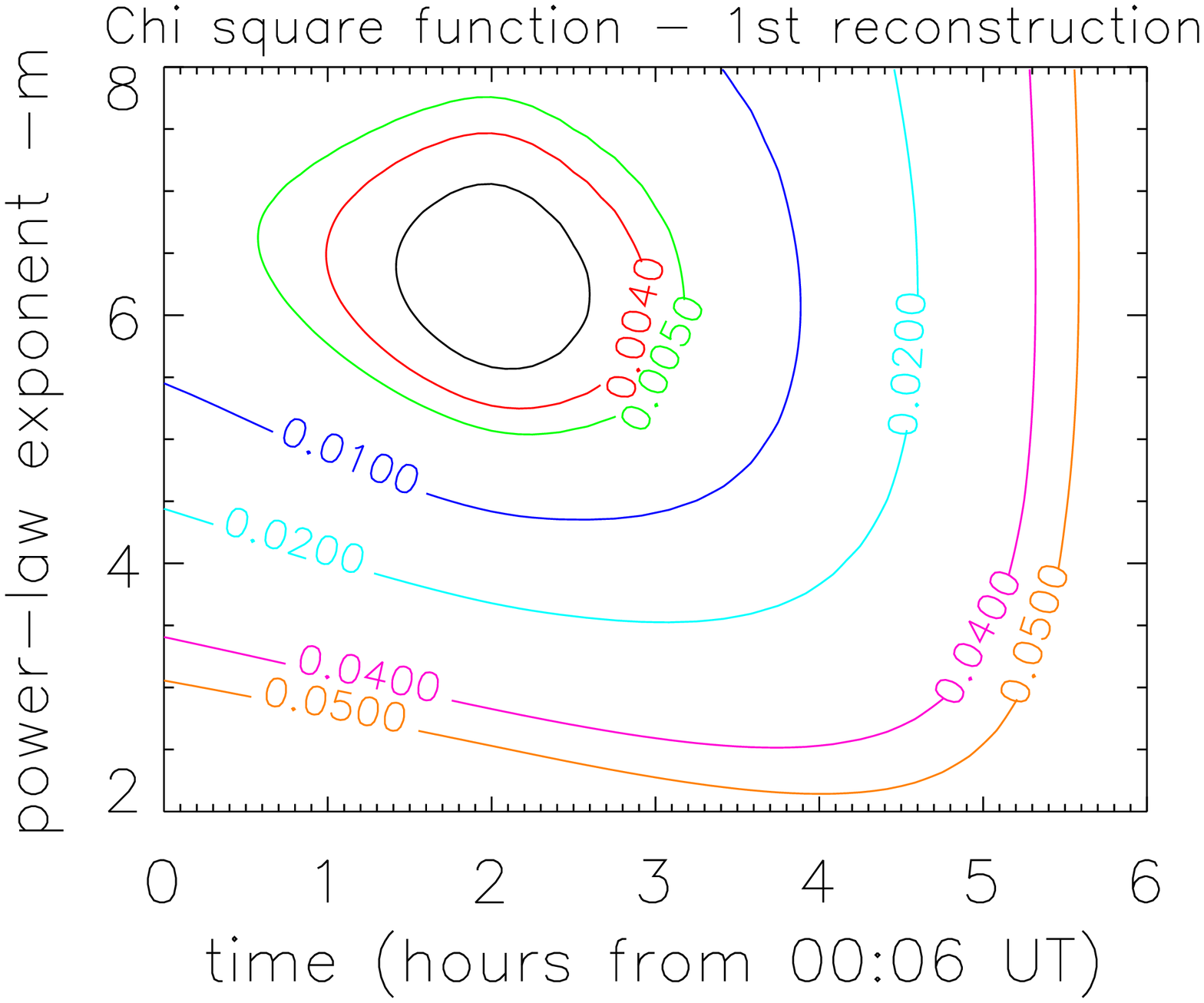}
      \includegraphics[width=.49\textwidth,type=eps,ext=.eps,read=.eps]{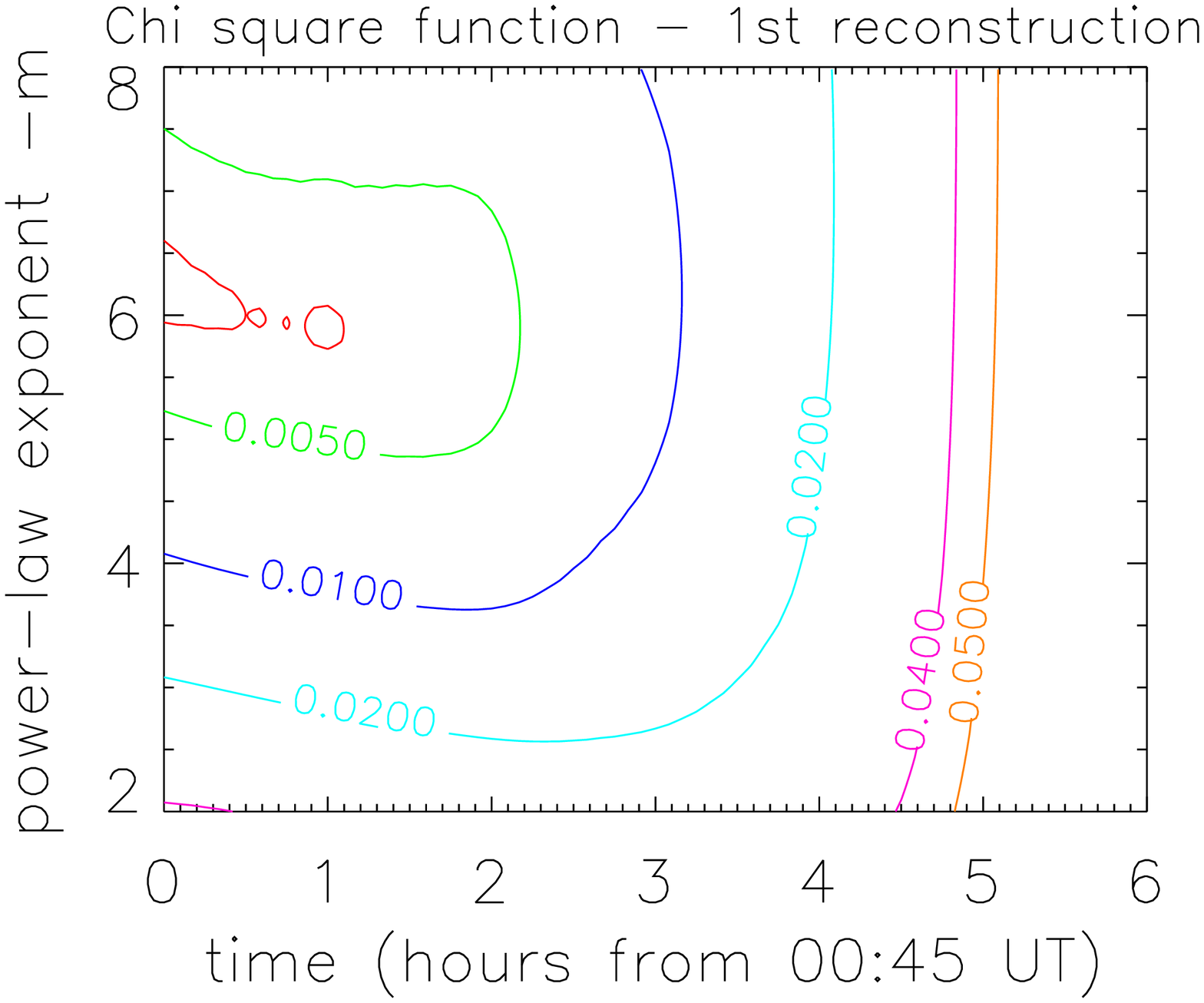}
      \caption{Contours of $\chi^2$ as function of $m$ and $t_0$. Left panel: the contours derived from the EUVI-A and -B 304~\AA\ reconstruction. Right panel: the contours derived from the SWAP 174~\AA\ and EUVI-B 304~\AA\ reconstructed points.}
    \label{fits}
 \end{figure*}

We found that depending on the selection of the initial time ($t_{0}$) of our fit, the fitting procedure returns different values of the exponent $m$. To select the values of $t_{0}$ and $m$ that best fit our data, we employ the following procedure:
we minimize the function  $\chi^{2} = \sum\left( H\left( t_{i} \right)-h\left( m,t_{i}-t_{0},h_{0},v_{0},a_{0}\right) \right)^2$ with respect to $h_{0}$, $v_{0}$, and $a_{0}$ for values of $m$ between 2 and 8 and $t_0$ between values corresponding to 00:00~UT and 10:00~UT. Here $H(t_i)$ is the reconstructed point at the time $t_i$. In Figure~\ref{fits}, we have plotted contours of $\chi^{2}$ as function of $m$ and $t_0$ to find the optimal parameters for the fit. The left panel of the figure shows the contours derived from our EUVI-A and -B reconstruction, while the right panel shows the contours derived from our SWAP and EUVI-B reconstructed points. The values of $m$ and $t_0$ that minimize $\chi^{2}$ are around $m=6$ and $t_0$ = 02:06~UT for the first reconstruction and around $m=6$ and $t_0$ = 01:57~UT for the second reconstruction. It is worth pointing out that the figure (right panel) also shows that, since $(t-t_0)^m$ changes very slowly for $t_0 < 1$ when $m$ is large, the procedure is not always very sensitive to $t_0$.

\subsection{Spline Smoothing}
The second method that we applied to the reconstructed points was a smoothing spline. For this we used the \emph{R}-based routine: \textsf{smooth.spline.r} (\emph{R} is a programming language for statistical computing.) In this case, the curve that we used to fit our data, $H(t)$, is represented by a cubic smoothing spline, $h_\mathrm{spl}(t)$.
The fitting method consists of minimizing the expression:
\begin{equation}
\frac{1}{n}\sum\limits_{i}{\frac{\left( H\left( t_i\right) -h_{spl}\left( t_i\right) \right)^2}{\sigma^2\left( t_i\right)}}+\lambda \int{\left( h_{spl}''\left( t\right) \right)^2\mathrm{dt}}.
\end{equation}
Here $H(t_i)$ is the reconstructed point at the time $t_i$, $h_\mathrm{spl}(t_i)$ is a cubic spline determined from Equation (2), $h_{spl}''(t)$ is its second derivative, $\sigma (t_i)$ is the measured error of the data point at $t_i$, and $\lambda$ is the so-called smoothing parameter (\textit{e.g.} \opencite{Craven78} for an explanation of polynomial smoothing splines). We have chosen nine to ten knot points to characterize the spline so that the spline sections were around 39 minutes each. These resulted in nine (second reconstruction) to ten (first reconstruction) degrees of freedom for each spline.

In order to determine $\lambda$, we used the ``leave-one-out'' cross-validation method. This method involves using a single observation from the sample data for validation, and using the remaining observations to minimize the expression above. Iterating this procedure using each data point in the series for validation once allows us to determine the best possible value of $\lambda$ for smoothing the data without over-smoothing. (See \opencite{Craven78} for additional details on how this procedure works.)

\begin{figure*}[!th]
  \centering
  \includegraphics[width=.49\textwidth,type=eps,ext=.eps,read=.eps]{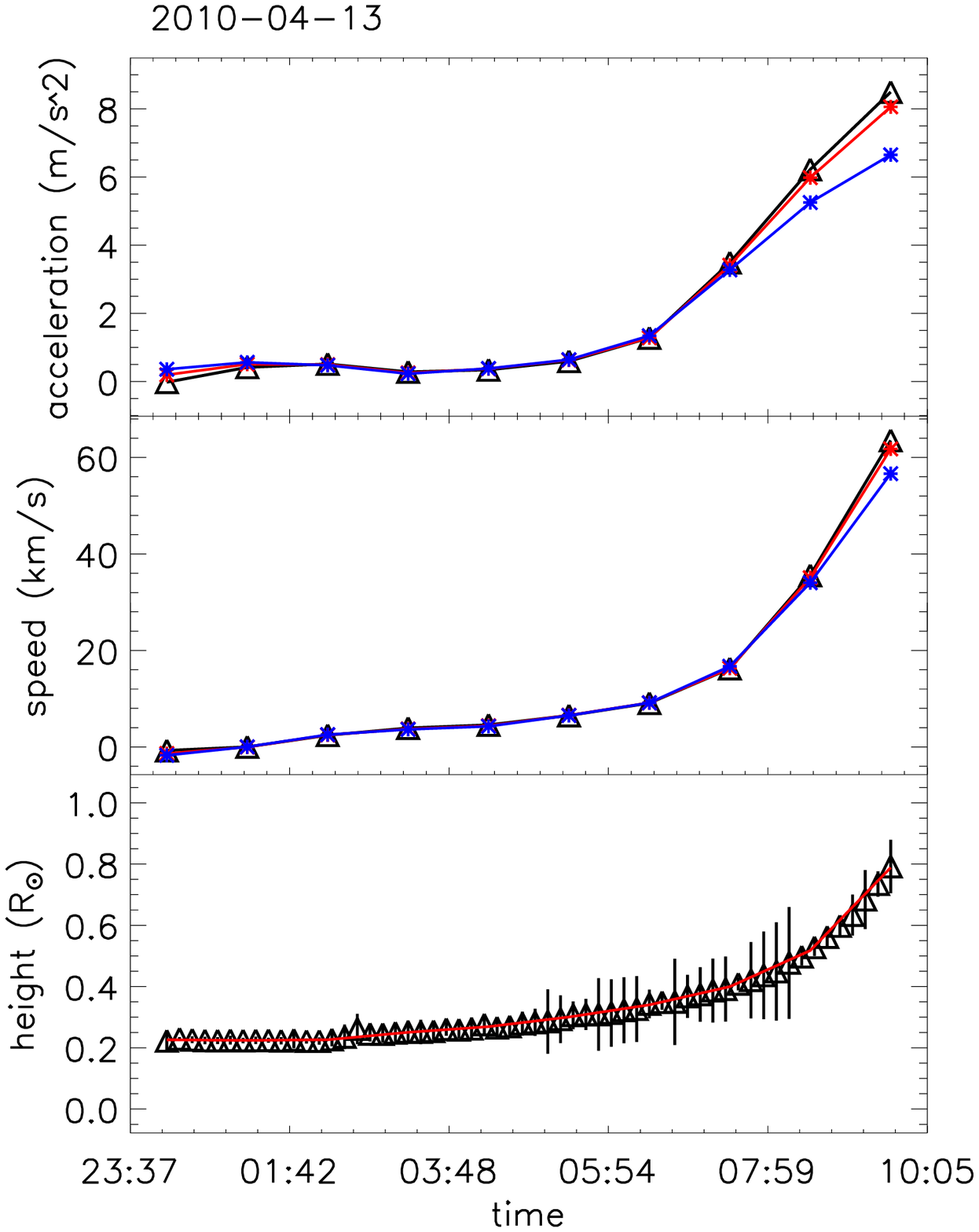}
  \includegraphics[width=.49\textwidth,type=eps,ext=.eps,read=.eps]{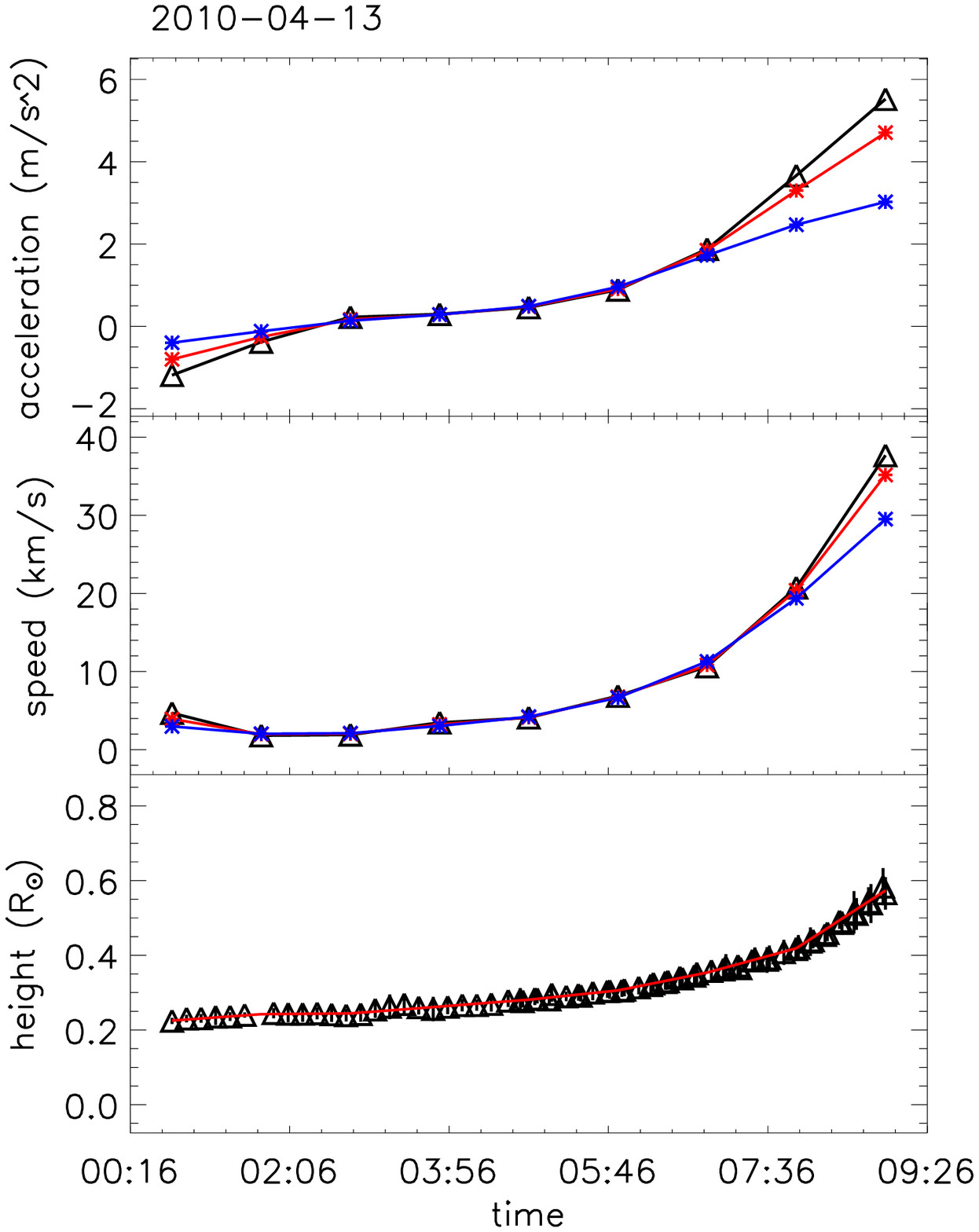}
  \caption{Lower panels: 3D height (above surface) \textit{versus} time (black symbols). The red curve represents the fitting of a cubic smoothing spline to the reconstructed data. Middle and upper panels: Speeds and accelerations derived from the smoothing procedure. The points are connected by lines for a better visualization. The left plot represents the points derived from the first reconstruction, with a smoothing parameter $\lambda = 2.62\times10^{-6}$ (black points), $1.38\times10^{-5}$ (red points) and $7.92\times10^{-5}$ (blue points)  and the right plot represents the points derived from the second reconstruction, with $\lambda = 2.17\times10^{-5}$ (black points), $1.1\times10^{-4}$ (red points) and $6.0\times10^{-4}$ (blue points).}
  \label{der_fits}
\end{figure*}

This method yields a value of $\lambda = 2.62\times 10^{-6}$ for the EUVI-A and -B reconstructed heights and a value of $\lambda = 2.17\times 10^{-5}$ for EUVI-B and SWAP reconstructed heights. These values were then used to produce a final fit incorporating all of the data points from which we were able to derive the speeds and accelerations (see Figure~\ref{der_fits}). 

In the figure we have also plotted the smoothed results for larger values of $\lambda$: $\lambda = 1.38\times 10^{-5}$ and $\lambda = 7.92\times 10^{-5}$ for the first reconstruction and $\lambda = 1.1\times 10^{-4}$ and $\lambda = 6.0\times 10^{-4}$ for the second reconstruction in red and green respectively. In general, larger values of $\lambda$ give a stronger weight to the second
term in Equation (2) so that its minimization leads to a smoother curve $h_{spl}$ and
fewer variations in the acceleration $h''_\mathrm{spl}$. At the ends of the
investigated time interval the lack of continuity constraints often leads
to a fit with reduced estimates of $|h''_\mathrm{spl}|$ if $\lambda$ is too large.
This effect can be seen in Figure 4 for the largest value of $\lambda$ chosen (blue curve). We therefore believe that the acceleration estimates for lower values of $\lambda$ (black and red curves) are more reliable.

\section{Results and Discussion}

\begin{figure*}[!th]
  \centering
  \includegraphics[width=.9\textwidth,type=eps,ext=.eps,read=.eps]{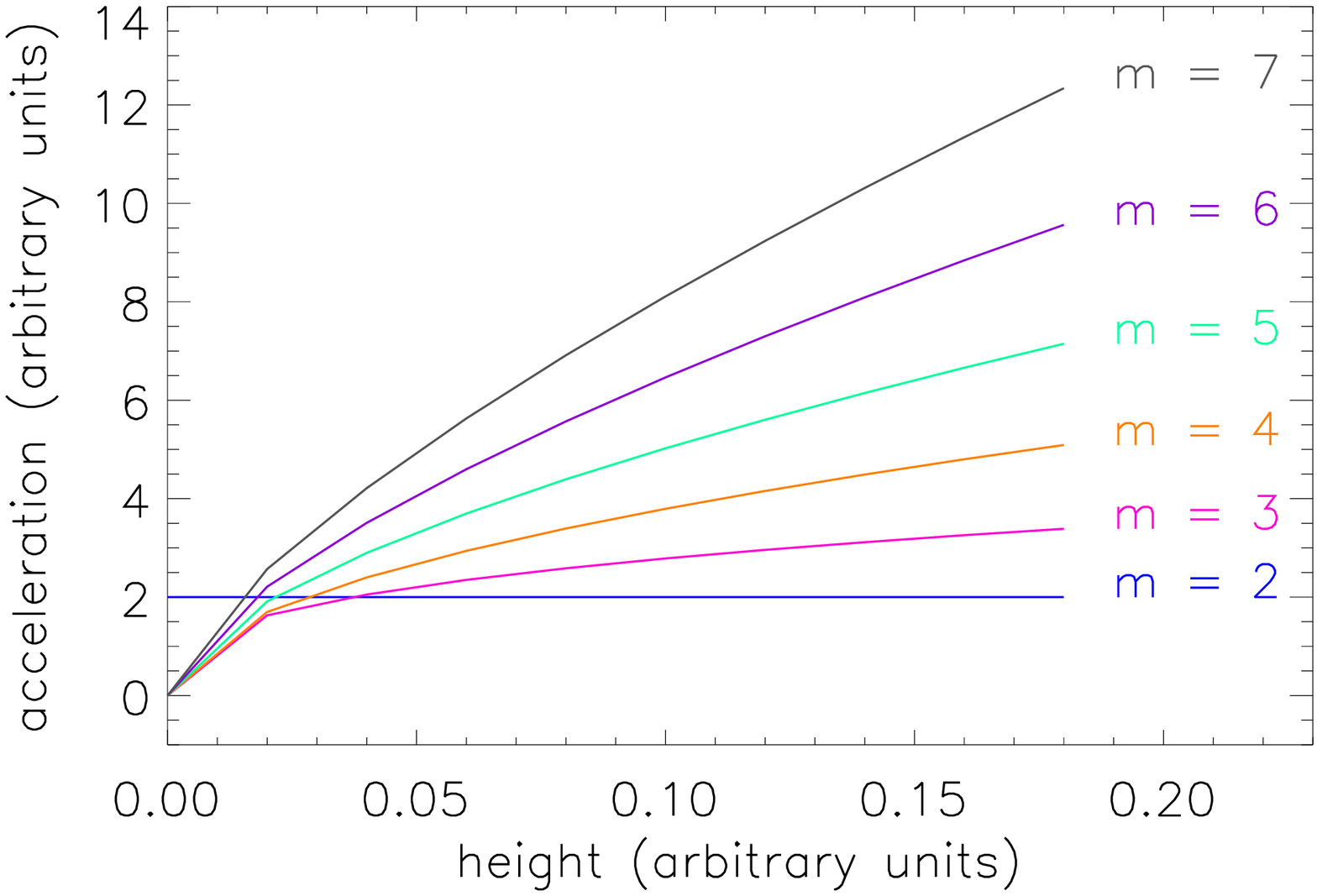}
  \caption{Schematic representation of acceleration \textit{versus} height for different values of $m$. The acceleration is derived from the power-law $h = t^m$ where $h$ represents the height (in arbitrary units) and $t$ is the corresponding time. The figure demonstrates that $m > 2$ in Equation (1) is equivalent to an increase of the acceleration with altitude.}
  \label{acc_fig}
\end{figure*}

\subsection{Results - Polynomial Fitting}
It is clear from Figure~\ref{fits} that the the optimal fit parameters for our function are $m \approx 6$ and $t_0 \approx \mathrm{02:00~UT}$. The value of $t_0$ tells us the time at which the forces that accelerated the eruption first set in, while the coefficient $m$ gives an indication of how immediate and impulsive those forces were. If $m = 2$, then the acceleration is constant. This can occur if the magnetic force ejecting the prominence is constant in time. For example, simulations of magnetic breakout indicate that the resulting movement can be well described by a constant acceleration profile \cite{Lynch04}. However, this does not seem to be the case in our event, as can be seen from Figure 4.

In Figure~\ref{acc_fig} we show a schematic representation of the acceleration as a function of height for polynomial fits with different values of $m$. For $m=6$, the acceleration increases smoothly, and the prominence continues to accelerate at greater heights as well. Compared with the results of \inlinecite{Schrijver08}, who found that values of $m$ between 2.7 and 3.8 better fit their observations, our value of $m$ is relatively large. This implies a smooth increase with height of the acceleration of the eruption, which suggests that the prominence was not accelerated by local reconnection, but rather was swept away as part of a large-scale relaxation of the coronal magnetic field.
The left panel of Figure~\ref{acc_height} shows the acceleration \textit{versus} height above the solar surface as derived from the polynomial fit to our data points. We see that starting at 0.3~solar radii, the acceleration begins smoothly increasing with height, while below 0.3 solar radii the acceleration is practically 0.

\subsection{Results - Splines}
\begin{figure*}[!th]
      \centering
      \includegraphics[width=.49\textwidth,type=eps,ext=.eps,read=.eps]{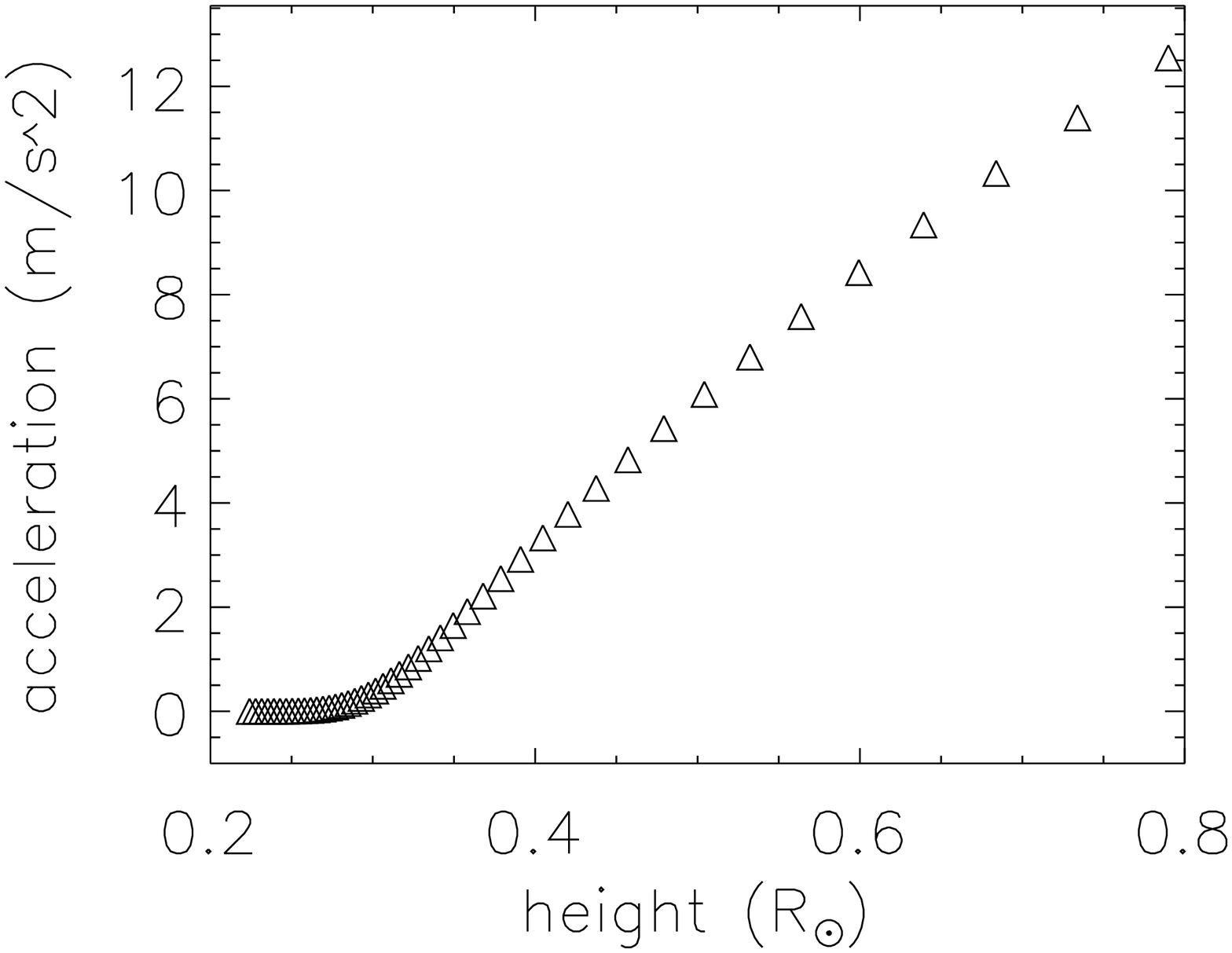}
      \includegraphics[width=.49\textwidth,type=eps,ext=.eps,read=.eps]{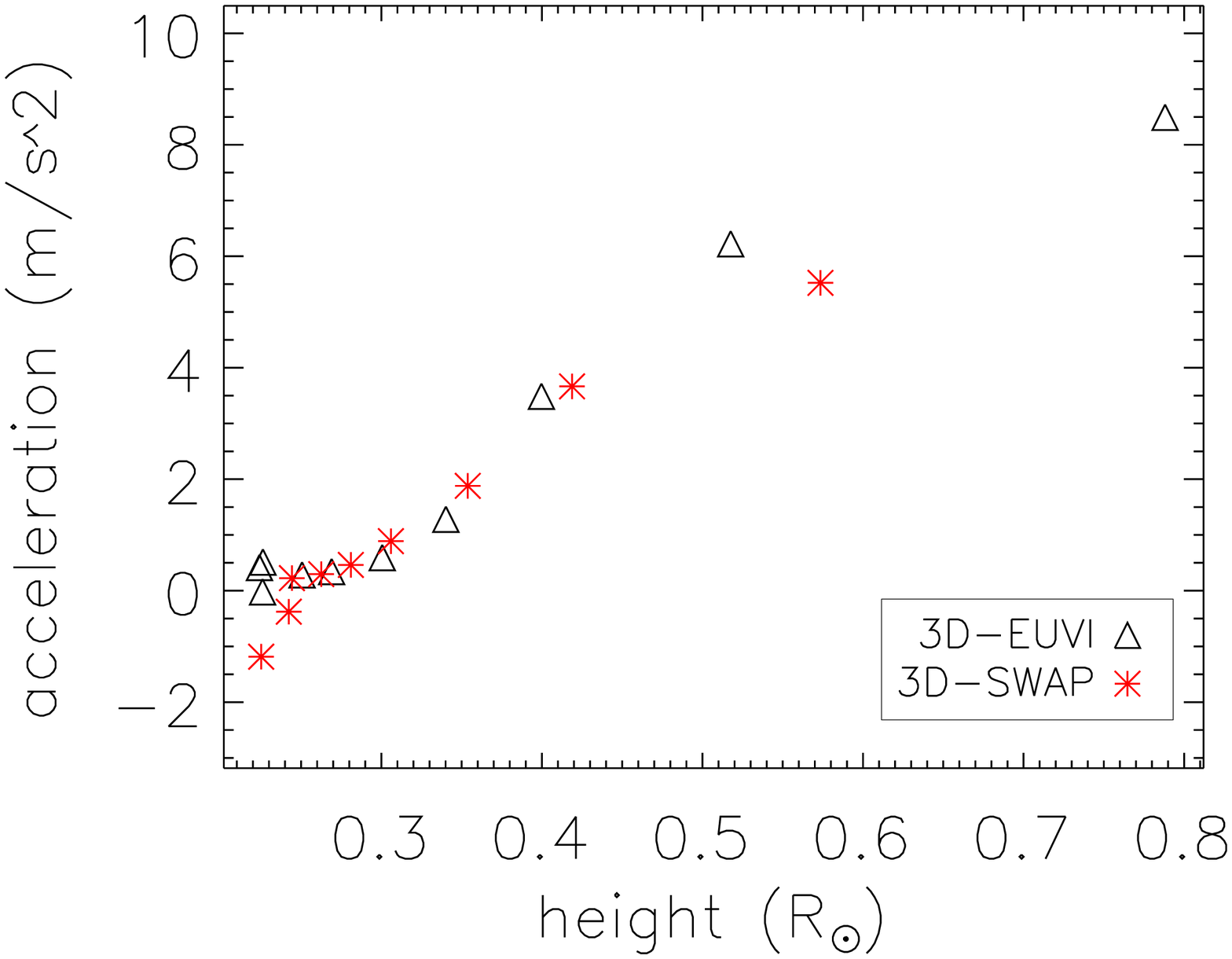}
      \caption{Left panel: Acceleration \textit{versus} height above solar surface as derived from the polynomial fit to the reconstructed points. Right panel: Acceleration versus height above solar surface as derived from smoothing spline procedure. Black triangles are the points from the first reconstruction and red asterisks are from the second reconstruction.}
    \label{acc_height}
 \end{figure*}

The right panel of Figure~\ref{acc_height} shows the acceleration derived from the spline smoothing procedure \textit{versus} the measured height of the prominence at each corresponding time. Here we see the acceleration is also smoothly increasing with height, beginning at a height close to 0.3 solar radii above the solar surface. (Before the prominence reaches this height, its velocity is essentially constant). Since the prominence reached this height around 06:15~UT, we conclude this was the end of the initiation phase and the beginning of the acceleration phase of the eruption. The acceleration phase continued until about 09:30~UT. Because of the limited fields-of-view of our observations, the propagation phase was not observed in our data.

\subsection{Discussion}
\inlinecite{Joshi11a}, \shortcite{Joshi11b} previously reported on this prominence eruption as well. They employed the triangulation method to generate a three-dimensional reconstruction of the complete prominence. They found that the prominence rises slowly for over five hours before erupting rapidly in the course of two hours (beginning at 08:36 UT). The longitudes that they derived from the reconstructed prominence span a large range, from 0 to 60~degrees, confirming that it was a very extended event. They also reported that when the prominence erupts, its axis twists in a clockwise direction, which they concluded was a consequence of the interplay of two motions: the overall non-radial motion of the prominence towards the equator and a helical twist of the axis of the prominence itself.

For different features they reconstructed along the prominence, they found an almost uniform motion (very small acceleration) during the slow-rise phase and a constant-acceleration behavior during the fast-erupting phase (acceleration ranging from 2 to 12~$\mathrm{m}$~$\mathrm{s}^{-2}$ for various tracked features).

\inlinecite{Joshi11b} also studied the late evolution of the eruption using images from the COR1 coronagraphs in STEREO. They found that the prominence continued to accelerate until it reached heights of about four solar radii, but the maximum CME acceleration occurred at heights less than two solar radii.

Our analysis, instead, was focused on the top of the prominence, which was clearly identifiable using images from three spacecraft: EUVI-A and -B on STEREO and SWAP on PROBA2. We have used two mathematical models to fit the HT diagram and to derive the acceleration. Namely, the m degree polynomial fitting (five degrees of freedom), and the spline smoothing (10 degrees of freedom). As reported in the literature, there are various methods to extract the acceleration from the HT profiles. In the usual case, a quadratic fit to the data is performed, either by using a three-point Lagrangian interpolation technique (\textit{e.g.} \opencite{Byrne09}, \opencite{Long08}, \opencite{Veronig08}) or by using the residual resampling bootstraping technique \cite{Byrne10,Long11}. The bootstraping technique consists in randomizing the residuals (\textit{i.e.}, the difference between the data and the fitted values), applying them to the original fit, refitting, and repeating this process a large number of times ($\approx$ 10 000 times). This technique proved to be statistically rigorous and produces a more accurate result than a simple model-fit to the given data \cite{Long11}. \inlinecite{Patsourakos10} have used a cosh function (described in \opencite{Sheeley07}) to fit their height--time profiles. The uncertainties in speed and acceleration were estimated by means of a Monte Carlo simulation. Their function could reproduce profiles ranging from nearly constant acceleration to impulsive acceleration. \inlinecite{Temmer10} have used a regularization method to minimize the errors in speed and acceleration, as derived from the height--time profiles. They compared the acceleration profiles derived from the regularization alghoritm with those derived from a least-square spline fit method. For sintetic data, without noise added, they found that the regularization algorithm represents the acceleration curve better than that from the spline fit. Nevertheless, for some of the real data (where the noise is intrinsic) both methods failed to reproduce the acceleration peak. However, the intrinsic errors in time derived from the regularization tool include the true solution. The overall shape of the acceleration profile (duration of acceleration) is well represented by both methods \cite{Temmer10}.

We used the classical least-square fitting to fit our data to the polynomial function. The second method makes use of a cubic smoothing spline procedure. The two mathematical models that we applied to the reconstructed heights both gave essentially the same result: the acceleration of the top of the prominence began smoothly and continues even as the prominence reaches greater heights in the corona. Although many observations suggest that eruptions unfold in two distinct phases, slow rise and rapid acceleration, in some events a rapid acceleration has not been observed. As we pointed out in the introduction, this suggests that factors other than impulsive reconnection may be responsible for the acceleration of these CMEs \cite{Kahler88}.

Our results confirm that this sort of eruption indeed occurs in the corona. We observed acceleration gradually increasing out to between 0.3 and 0.8~solar radii above the solar surface (see Figure~\ref{acc_height}). Such a gradual acceleration is unlikely to be caused by the kind of strong reconnection observed in the impulsive eruptions associated with solar flares, and is more likely the result of a magnetic field relaxing to a new equilibrium.

The values of acceleration derived in this study compare with those calculated by \inlinecite{Joshi11b}, but as opposed to their findings, we find that the acceleration is continuously increasing with height. This difference may be due to different features that they follow compared with us.

\section{Summary}
The 3D height--time diagram of an erupting prominence on 13~April~2010 was fitted with two mathematical models (the first characterized by five degrees of freedom and the second characterized by ten degrees of freedom). The results derived from the two methods show that the acceleration increases smoothly and continuously with height. This suggests that the prominence is not accelerated immediately by local reconnection but it is rather swept away as part of a large-scale relaxation of the coronal magnetic field to a new equilibrium.

We also found that two phases characterize the early evolution of the eruption: the initiation phase and the acceleration phase. By applying the two methods we were able to determine the exact moment of the two phases observed: the initiation phase characterized by a slow rise (constant speed) lasts from around 02:00 UT (0.2 solar radii above solar surface) until 06:15 UT (0.3 solar radii above solar surface) when the acceleration phase begins. The propagation phase was not observed in our data.
 

 \begin{acks}
M.M would like to acknowledge S. Patsourakos for providing the program on spline smoothing and for productive discussions in using this program. She would like to thank also W. Thompson and N. Rich for providing the programs on EUVI pointing correction and height--time measurements. The authors also thank B. Kliem for helpful discussions of our analysis of this event.
M.M. thanks MPS for financial support in carrying out this work. Part of her work was also covered from the project TE 73/11.08.2010. We acknowledge the use of SWAP/PROBA2 and SECCHI/STEREO data. SWAP is a project of the Centre Spatial de Li$\grave{e}$ge and the Royal Observatory of Belgium funded by the Belgian Federal Science Policy Office (BELSPO). The SECCHI data used here were produced by an International Consortium of the Naval Research Laboratory (USA), Lockheed--Martin Solar and Astrophysics Lab (USA), NASA Goddard Space Flight Center (USA), Rutherford Appleton Laboratory (UK), University of Birmingham (UK), Max-Planck Institute for Solar System Research (Germany), Centre Spatiale de Li$\grave{e}$ge (Belgium), Institut d'Optique Theorique et Appliqu$\acute{e}$e (France), Institut d'Astrophysique Spatiale (France).

 \end{acks}

%
%
\bibliography{bibliography.bib} 
\bibliographystyle{spr-mp-sola-cnd}

%
%

\end{article} 
\end{document}